\documentclass[superscriptaddress,twoside,twocolumn,bibnotes,tightenlines,aps,floats,showpacs,floatfix]{revtex4}
\usepackage{times}
\usepackage{dcolumn}
\usepackage{graphicx}
\usepackage{mathptm}
\usepackage{amsmath}

\begin{document}
\graphicspath{{.}{./Figures/}}
\title{Resonance tongues and patterns in periodically forced
reaction-diffusion systems}
\author{Anna L. Lin}
\affiliation{Center for Nonlinear and Complex Systems
and Department of Physics, Duke University, Durham, NC 27708}
\author{Aric Hagberg} \email{aric@lanl.gov}
\affiliation{Mathematical Modeling and Analysis, Theoretical Division,
Los Alamos National Laboratory, Los Alamos, NM 87545}
\author{Ehud Meron} \email{ehud@bgumail.bgu.ac.il}
\affiliation{Department of Solar Energy and Environmental Physics,
BIDR, Ben-Gurion University, Sede Boker Campus 84990, Israel}
\affiliation{Physics Department, Ben-Gurion University, Beer Sheva
84105, Israel}
\author{Harry L. Swinney}
\affiliation{Center for Nonlinear Dynamics and Department of Physics,
The University of Texas at Austin, Austin, TX 78712}

\begin{abstract}
Various resonant and near-resonant patterns form in a
light-sensitive Belousov-Zhabotinsky (BZ) reaction in response to a
spatially-homogeneous time-periodic perturbation with light.
The regions (tongues) in the forcing frequency and forcing amplitude
parameter plane where resonant patterns form are identified
through analysis of the temporal response of the patterns.
Resonant and near-resonant responses are distinguished.  The
unforced BZ reaction shows both spatially-uniform oscillations
and rotating spiral waves, while the forced system shows patterns such
as standing-wave labyrinths and rotating spiral waves.
The patterns depend on the amplitude and frequency of the perturbation,
and also on whether the system responds to the forcing near the
uniform oscillation frequency or the spiral wave frequency.
Numerical simulations of a forced FitzHugh-Nagumo
reaction-diffusion model show both resonant and near-resonant
patterns similar to the BZ chemical system.  

\end{abstract}

\date{\today}
\maketitle

\section{Introduction}

An oscillator forced by a periodic external perturbation entrains
to the forcing for certain values of the perturbation frequency
and amplitude.  This behavior is observed in a wide range of
biological, chemical and physical systems, for example, in
circadian rhythms such as the sleep-wake cycle forced
by the sun~\cite{Glas:01}, in the tips of chemical spiral
waves forced with light~\cite{BrEn:93,SZMu:93,MaBa:96,HaKa:99,KaZy:99},
and in arrays of Josephson's junctions~\cite{BBJe:84}.

The entrainment to the forcing can take place even when the
oscillator is detuned from an exact
resonance~\cite{Arnold:83,JBB:84,Krau:95}. In this case, a
periodic force with a frequency $f_f$ shifts the oscillator from
its natural frequency, $f_0$, to a new frequency, $f_r$, such that
$f_f/f_r$ is a rational number $m\mbox{:}n$. When the forcing amplitude is
too weak this frequency adjustment or locking does not occur; the
ratio $f_f/f_r$ is irrational and the oscillations are
quasi-periodic. In dissipative systems frequency locking is the
major signature of resonant response. Nearly conservative systems
show in addition a large increase in the amplitude of
oscillations.

The response of a two-dimensional array of coupled nonlinear
oscillators, or of a two-dimensional oscillating field is much
less well understood.  For a periodically forced single
oscillator, the structure in the parameter plane of the forcing
frequency $f_f$ and amplitude $I$ contains many universal features
identified with frequency locking~\cite{GlLi:88}, but it is
unclear if these features persist or change in the more complex
case of spatially extended systems where patterns can form.
Patterns resulting from time-periodic forcing
~\cite{POS:97,LBMS:00,lin00:_four_phase,MLKS:02,yochelis02:_devel,Denn:00,DZEp:01,GSMT:00,GWST:01,GCKa:98,HeKa:01},
spatially periodic forcing~\cite{DBZE:01}, and global
feedback~\cite{VYDZE:00,VZEp:01,ZyEn:02} have been studied in the
past, but a whole phase diagram showing a resonance
structure for spatially extended systems
has not been reported. Moreover, it is not even clear to what
extent the familiar concept of resonance applies to spatially
extended systems.

To investigate a phase diagram showing multiple resonances in
spatially extended systems, we applied periodic perturbations to a
light sensitive form of the Belousov-Zhabotinsky (BZ) chemical
reaction-diffusion system and measured the temporal response and
pattern formation.
In the course of this investigation a new mode of response has
been identified.
Patterns may fail to lock to the forcing frequency but still
respond by showing an $m$-peaked distribution of the 
oscillation-phase as in resonant patterns.
We refer to this response mode as ``near-resonant.''

\begin{figure*}[htb]
  \centering\includegraphics[width=6.0in]{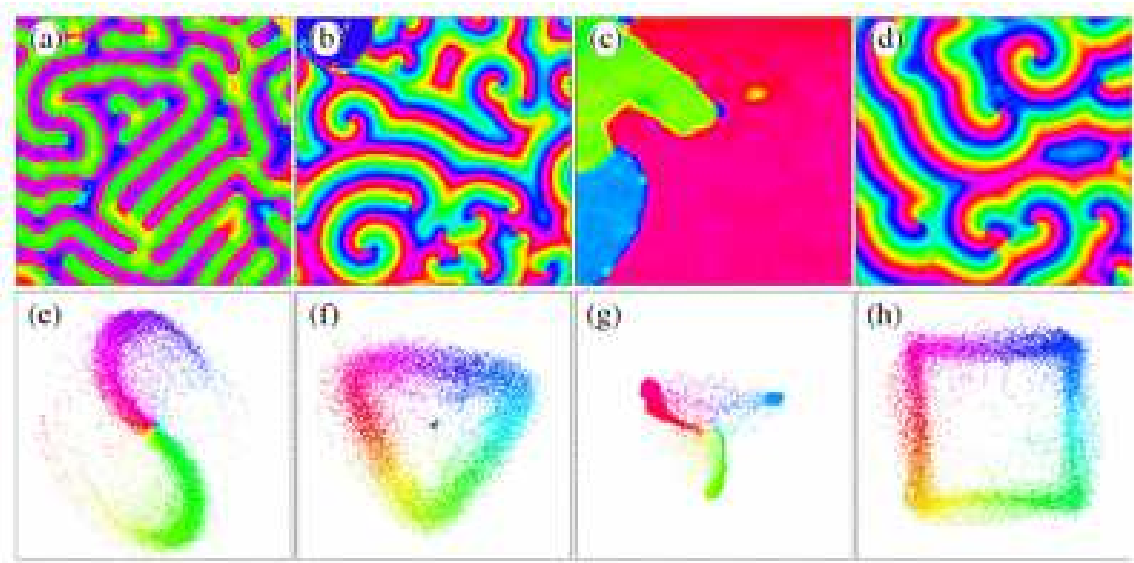}
  \caption{(Color online)
    Patterns in the periodically forced BZ reaction.
    (a)-(d) show a $11.5\times11.5$ $\rm mm^2$ size
    region of patterns formed at different forcing
    frequencies and amplitudes.  The data are processed to
    show the pattern only near the sub-harmonic response frequency.
    Patterns (a) and (c) are
    frequency-locked (resonant).
    Patterns (b) and (d) are near-resonant but not
    frequency-locked.  The frames (e)-(h) show the same data represented
    in the complex phase plane, see~\cite{LBMS:00}.  The colors in the phase
    plane vary with the angle $[0,2\pi]$ and are the key
    to the pattern images above. See the text for details.
    (a,e) 2:1 two phase standing-wave pattern;
    (b,f) 3:1 three phase rotating spiral
    (c,g) 3:1 three phase standing-wave pattern;
    (d,h) 4:1 four phase rotating spiral.
    Chemical conditions are given in~\protect{\cite{BZ1}}.
  }
  \label{fig:many}
\end{figure*}

In addition to the nonuniform distribution of the oscillation
phase, resonant and near-resonant, patterns can also
be characterized by the {\em shape} of the phase in the complex phase plane.
The phase of unforced spirals has a circular shape in the
complex phase plane but forcing breaks the circular symmetry.
At high enough forcing this is visible as a $m$-fold symmetry
in the phase plane.
Examples of $m\mbox{:}1$ patterns observed in the BZ system, where
$m\mbox{:}1$=$f_f/f_0$ and $m=2,3,4$ are shown in Fig~\ref{fig:many}.  Unlike
the single oscillator case, a spatially extended system can exhibit
phase waves and other phase patterns.  In Fig.~\ref{fig:many} each
pattern is shown in two representations: in the real space $x-y$
plane, and in the complex phase plane. 

In this paper we construct an experimental phase diagram in the
forcing frequency and amplitude parameter plane of resonant and 
nearly resonant $m\mbox{:}n$
responses and identify the pattern types that lead to the two
responses.  The experimental set-up and determination of resonance
tongues are described in Sec.~\ref{sec:bzexp} and
Sec.~\ref{sec:tongues}, respectively. The qualitatively different
patterns observed in the experiments are presented in
Sec.~\ref{sec:patts}.  A forced reaction-diffusion model, modified
FitzHugh-Nagumo equations, is introduced in Sec.~\ref{sec:fhn}, which
is followed by a discussion of the model and the insight it provides
into the mechanisms of pattern formation in the experimental system.
A general discussion and summary of the results are given in
Sec.~\ref{sec:discussion}.

\section{Belousov Zhabotinsky Chemical System}
\label{sec:bzexp}

The oscillatory chemical reaction occurs in a thin, porous-glass
membrane (0.4 mm thick, 25 mm in diameter), which is in contact on
each side with continuously refreshed reservoirs of reagents for the
ruthenium-catalyzed Belousov-Zhabotinksy (BZ) reaction~\cite{diagram}.
Each reservoir is well-stirred and the reagents diffuse from them into
the membrane where they react.  The chemical concentrations in the
reservoirs are given in Ref.~\cite{BZ1,BZ2}.  Visualization of the
patterns is achieved using a low intensity tungsten lamp,
which measures the optical density of the concentration patterns
in the membrane without affecting the chemical reactions.

\begin{figure}
  \includegraphics[width=1.70in]{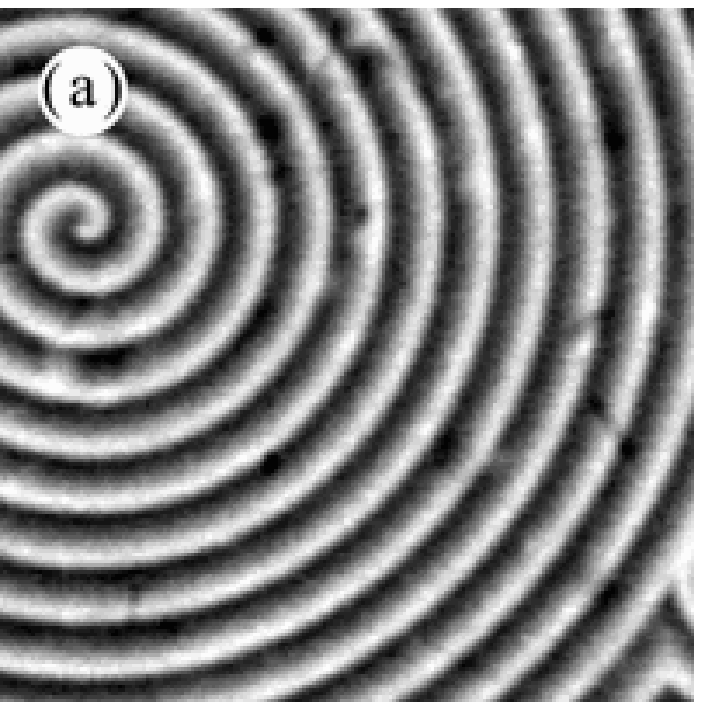}\includegraphics[width=1.70in]{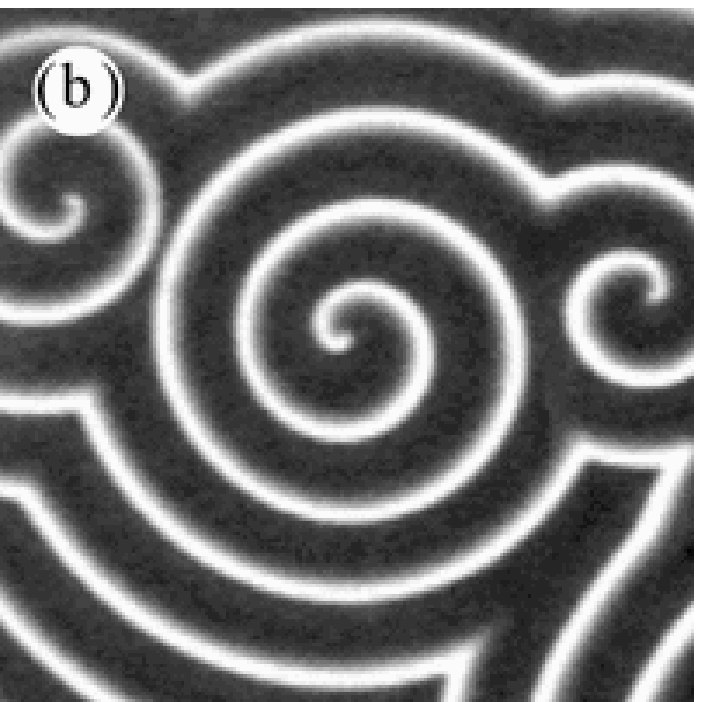}
  \caption{
    Unforced spiral patterns for the two sets of chemical conditions
    used for the experiments presented.
    (a) Spiral waves with a shorter period,
    $[BrMA]$=0.220 M, $[BrO_3^-]$=0.230 M~\cite{BZ1}.
    (b) Spiral waves with a longer period, $[BrMA]$=0.300 M,
    $[BrO_3^-]$=0.136 M~\cite{BZ2}.  The
    images show a $9$ mm $\times$ $9$ mm region of the
    pattern.  Dark regions correspond to high concentrations of Ru(II).
    Taken from~\protect\cite{MLKS:02}.
    All data reported in this paper are taken under
    the conditions for the spirals on the left, except for
    Fig.~\protect\ref{fig:alltongues}(b).
  }
  \label{fig:spirals}
\end{figure}

For the chemical concentrations used in the present experiments the
unforced system exhibits rotating spiral patterns.  The patterns are
sustained indefinitely in time because the reaction products leave the
membrane by diffusion into the reservoirs, and reservoir
concentrations are maintained by continuous feeds.  We used two
different sets of chemical conditions~\cite{BZ1,BZ2}, one creating
spirals with a higher frequency ($f_s=0.072~\mbox{Hz}$) and one
creating spirals of a lower frequency ($f_s=0.020~\mbox{Hz}$).
Examples of the unforced spiral waves for both sets of chemical
conditions are shown in Fig.~\ref{fig:spirals}.

In addition to the spiral frequency, the BZ system has another natural
frequency: the unforced spatially-homogeneous oscillation frequency.
Since perturbations always lead ultimately to the formation of spiral
waves in the membrane, we determined the homogeneous oscillation
frequency $f_0$ by the following method.  The membrane was exposed to
a spatially uniform high-intensity pulse of light for 30 s, which
resets the system so the entire membrane is oscillating with the same
frequency and phase.  The frequency of this spatially
uniform oscillation, determined by the chemical kinetics, was found in
our previous work to be essentially independent of the chemical
concentrations used in our study, $f_0=0.020~\mbox{Hz}$~\cite{MLKS:02}.
The uniform oscillations eventually evolve to
rotating spiral waves which fill the system.  However, the frequency
of the spiral waves {\sl does} depend on the chemical conditions.  The
ramifications of this dependence will be discussed in the next
Section.

\section{Multiple resonance tongues}
\label{sec:tongues}

The chemical reaction is forced by illuminating it with spatially
homogeneous light that is periodically
blocked~\cite{LBMS:00,lin00:_four_phase,MLKS:02}; the durations of the
illuminated and blocked portions of each cycle are equal, i.e., the
intensity modulation is a square wave.  To investigate the temporal
response of the patterns to the forcing we varied the intensity $I$
and frequency of the periodic light forcing $f_f$.  The light intensity $I$
is the control parameter for the forcing amplitude.
We examined the temporal response to determine which, if any,
tongue the pattern belonged to, and examined
the existence, shape, and ordering of the resonance tongues (see
Sec.~\ref{sec:tongues}).

\subsection{Determining Temporal Resonance}
\label{sec:temp}

\begin{figure}
\includegraphics[width=3.25in]{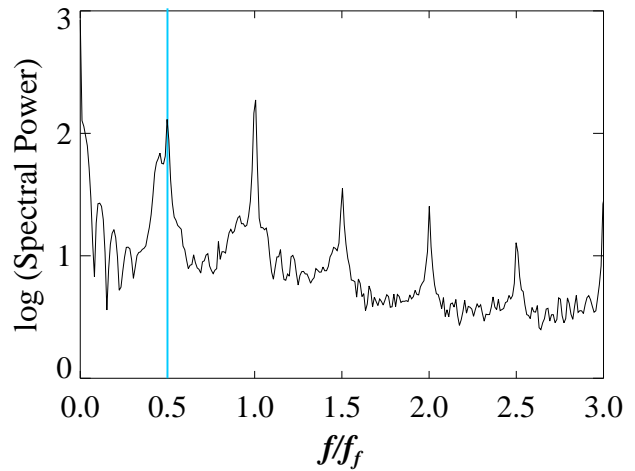}
\caption{Averaged power spectrum for a 2:1 resonant labyrinthine pattern.
The response frequency has been normalized to the forcing frequency $f_f$.
The peak at 0.5 is the subharmonic response to the forcing.  The
large peak at the forcing frequency ($f=f_f$) is due to our imaging
method, which captures part of the forcing light output.
In this paper we only consider responses subharmonic
to the forcing frequency ($f/f_f < 1)$).
}
\label{fig:power}
\end{figure}

The experimental data were collected as a time-sequence of pattern
snapshots.  The natural oscillation period of the reaction for the
conditions used was about $50$ s.  Typically, images were recorded
every $2$ s for one hour and a central $240 \times 240$ pixel (23 mm
$\times$ 23 mm) region of the pattern was analyzed.  The Fourier
transform of the time series for each pixel was calculated to obtain
an average power spectrum for the pattern.  Figure~\ref{fig:power}
shows a typical averaged power spectrum for a resonant pattern when
the forcing frequency $f_f$ was twice the uniform oscillation frequency
$f_0$. The largest subharmonic frequency peak appears at $f=f_f/2$, as
indicated by the vertical line.

\subsection{Tongues}
\begin{figure}
\centering
\includegraphics[width=3.25in]{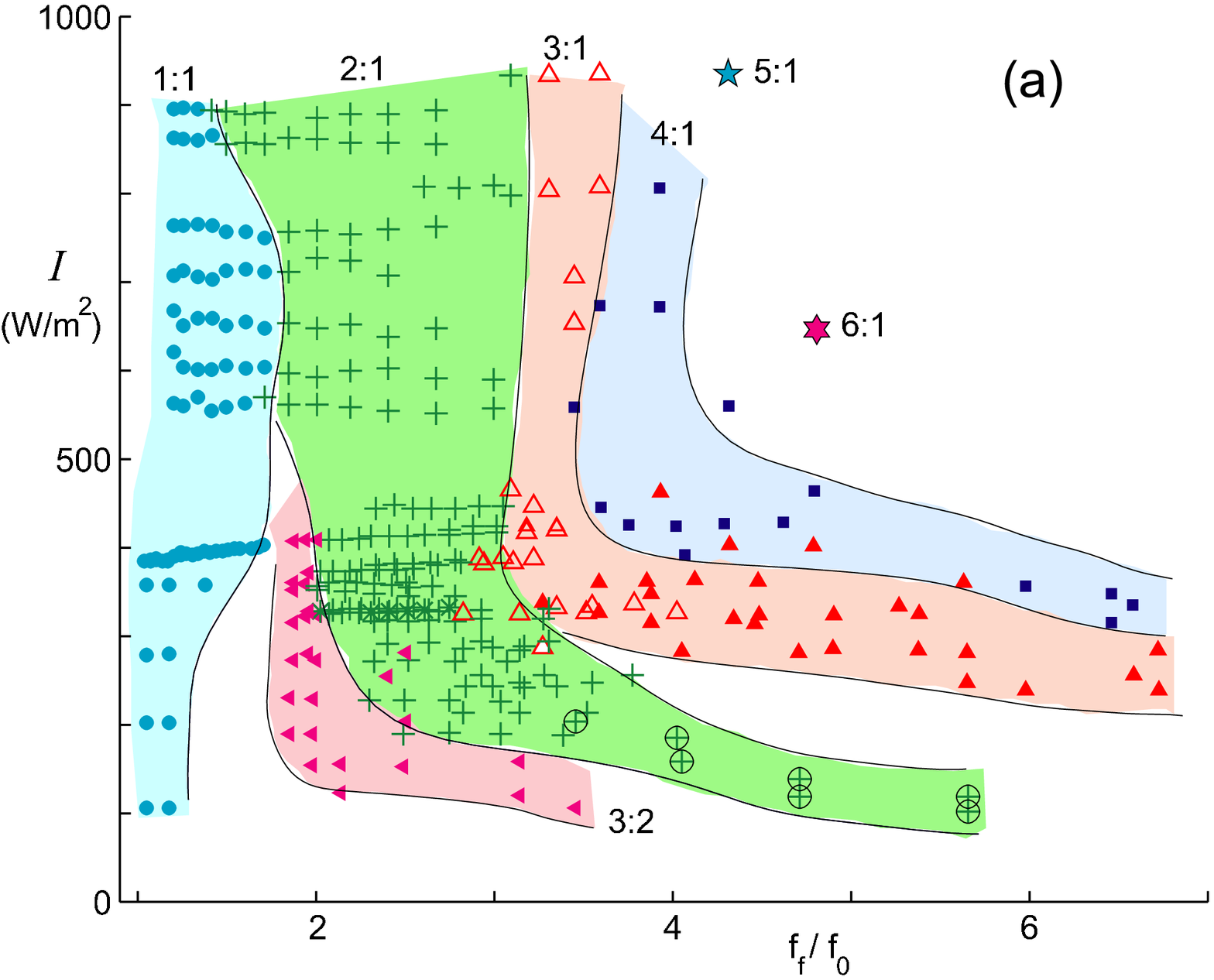}
\includegraphics[width=3.25in]{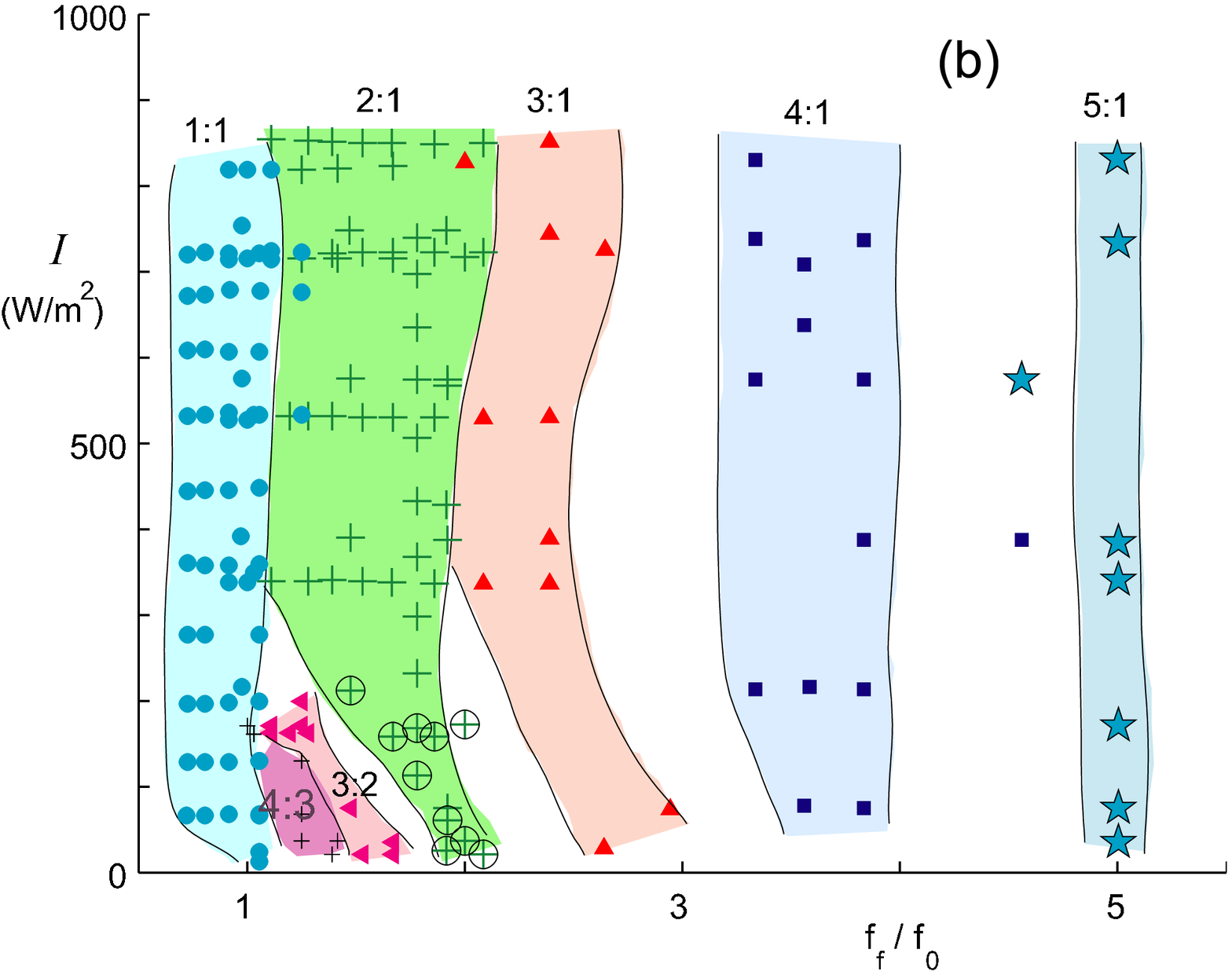}
\caption{(Color online)
  The largest $m\mbox{:}n$ tongues observed in the frequency-intensity plane of
  the spatially extended BZ system. Chemical conditions are those used
  in Fig.~\protect\ref{fig:spirals}(a)~\cite{BZ1}.  
  (b) The largest $m\mbox{:}n$ tongues
  observed for the chemical conditions used
  in Fig.~\protect\ref{fig:spirals}(b)~\cite{BZ2}.  
  The homogeneous frequency in
  both cases is $f_0=0.020$~Hz while the spiral frequency is 
  (a) 0.072~Hz,
  (b) 0.020~Hz.  
  Each symbol type represents a different $m\mbox{:}n$ response.  The
  patterns (points) within the solid curves respond sub-harmonically
  with the forcing frequency.  The bottom plot is taken from
  \protect{\cite{Martinez_MS}}.
}
\label{fig:alltongues}
\end{figure}

Using the method described in the previous section, we found tongues
in the forcing parameter space where the pattern responds at or near
$m\mbox{:}n$ resonances.  We obtained a phase diagram for each of the two
chemical conditions~\cite{BZ1,BZ2}, shown in
Fig.~\ref{fig:alltongues}(a),(b).  If the peak of the strongest mode
subharmonic to the forcing was within $\pm 3\%$ of the forcing
frequency, we considered the pattern to be responding to the forcing,
and it is included within an $m\mbox{:}n$ resonance region in
Fig.~\ref{fig:alltongues}.  This criterion is consistent
with the observation of $m$-peaked distributions of the oscillation
phase.  Some of the patterns meeting this
criterion are resonant while others are near-resonant, i.e., they are
quasi-periodic patterns with an $m$-peaked phase distribution.

We varied the forcing frequency and intensity in the experiments and
explored the temporal resonant response as we moved through the
parameter space, and the results are shown in
Fig.~\ref{fig:alltongues}.  Each symbol type represents a different
$m\mbox{:}n$ resonance.  The curves in Fig.~\ref{fig:alltongues} are drawn to
guide the eye to the tongues in the $f_f$-$I$ plane with different $m\mbox{:}n$
responses.  Only the largest resonance tongues are plotted.
In addition to the $m\mbox{:}1$ tongues (and the 4:3 and
3:2 tongues) shown, we observed several higher order $m\mbox{:}n$ states (e.g.,
5:7, 5:1, 6:1, 10:1), which spanned control parameter ranges too
narrow to be maintained.  In all cases the different $m\mbox{:}n$ tongues were
ordered in a Farey sequence, similar to the Devil's staircase ordering
of resonance tongues for two coupled oscillators~\cite{PBak:86} and
for the homogeneous BZ reaction~\cite{maselko86, maselko87}.

We investigated two different chemical conditions.  The chemical
conditions that yield 0.072~Hz spirals have $m\mbox{:}n$ tongues that bend
toward higher frequency as the light intensity $I$ is decreased
[Fig.~\ref{fig:alltongues}(a)].  For the chemical conditions that
yield 0.020~Hz spirals, the tongues do not bend much at low frequency
[Fig.~\ref{fig:alltongues}(b)].  The bending of the tongues is caused
by a shifting from $f_0$ = 0.02~Hz resonance at high forcing intensity
(the uniform oscillation frequency) to the near-resonant response of
the spiral wave frequency $f_s$ for lower forcing intensity.  Since
the spiral wave frequency for the data in Fig.~\ref{fig:alltongues}(b)
is the same as the uniform oscillation frequency, the tongues do not
bend in that case~\cite{MLKS:02,Martinez_MS}.

\subsection{A quantitative measure of patterns}

Spatial Fourier transforms and correlation functions do not capture
the temporal aspects of the patterns and did not differentiate the
data well because the patterns were often comprised of multiple
wavelengths and orientations.  Therefore, instead of computing spatial
Fourier transforms, we analyzed the temporal Fourier transform
calculated for each point in the
pattern~\cite{LBMS:00,lin00:_four_phase,MLKS:02}.  The power spectrum of the
signal, averaged over the spatial pattern, gives information about the
strongest frequency response.  Additionally, the data were filtered to
keep the strongest response and then inverse Fourier transformed.  The
complex amplitudes of the filtered system give information about the
phase distribution of the pattern.  Qualitatively different patterns
were found to exhibit different shapes in the complex phase-plane
representations~\cite{LBSA:00} [see Fig.~\ref{fig:many}(e-h)].

\section{Pattern formation}
\label{sec:patts}

We now describe the asymptotic patterns observed (after the decay of
transients) in the $m\mbox{:}1$ tongues for different forcing frequencies and
amplitudes.
The patterns can be divided into two categories: those which are
resonant with the forcing and those which are near-resonant. 
The resonant patterns are standing waves which lock to the
forcing frequency and show $m$ peaks in the phase response.
The near-resonant patterns are traveling waves and spiral waves which do not
lock to the forcing but still show $m$ peaks in the phase response.

\begin{figure}
  \includegraphics[width=3.0in]{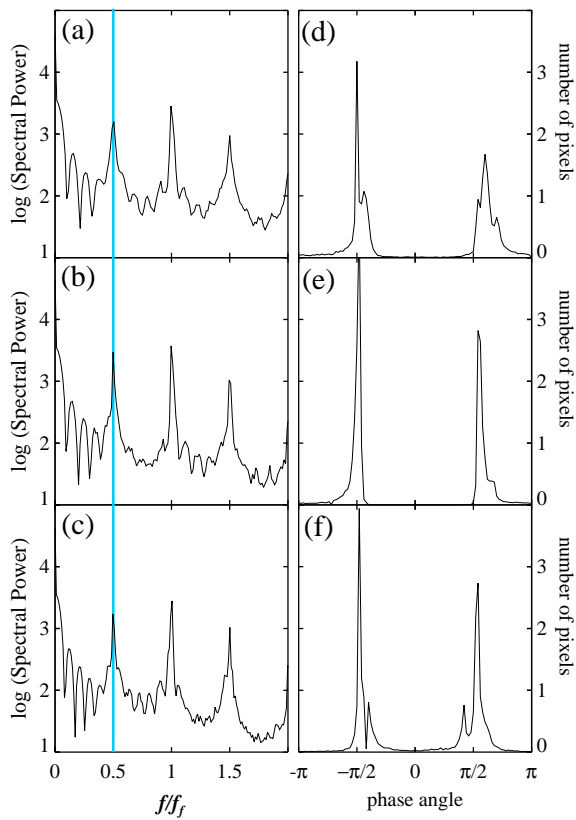}
  \caption{
    Power spectra and phase histograms for resonant 2:1 standing-wave patterns.
    Frames (a)-(c) show the power spectra of the response
    as the forcing frequency is varied across the tongue.  The power
    spectra, normalized to the forcing frequency $f_f$, show that
    the largest response is at exactly half the forcing frequency
    ($f/f_f=0.5$), as indicated by the vertical line.
    Frames (d)-(f) show histograms of the phase of the pattern
    near the peak response.  The histograms have two peaks corresponding
    to concentrations of the pattern in regions that are separated in
    phase by $\pi$.
    Forcing frequency:
    (a,d) $f_f=0.0333$~Hz,
    (b,e) $f_f=0.0357$~Hz,
    (c,f) $f_f=0.0416$~Hz.
  }
  \label{fig:bz-spiral}

\end{figure}

\begin{figure}
  \includegraphics[width=3.0in]{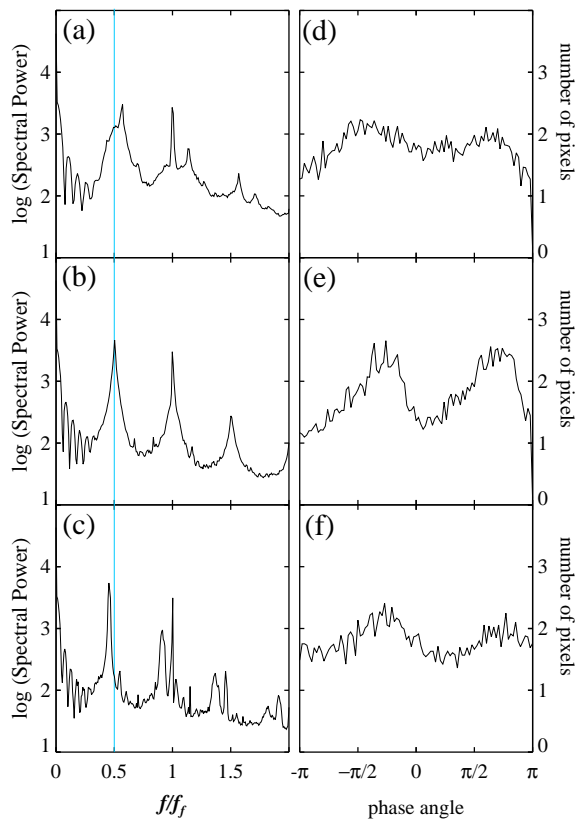}
  \caption{
    Power spectra and phase histograms for near-resonant 2:1
    spiral wave patterns.  Frames (a)-(c) show power spectra of the
    response as the forcing frequency is varied across the tongue.  The
    power spectra, normalized to the forcing frequency $f_f$, show that
    the largest response is nearly half (but not exactly half) the forcing
    frequency when the forcing is detuned from exact resonance
    ($f/f_f=0.5$ as indicated by the vertical line); thus the spiral wave
    pattern does not lock to the forcing frequency.  The peaks at $f/f_f$
    = 1 in (a) and (c) include power from the forcing light reflected from
    the reactor face into the camera.  Frames (d)-(f) show histograms of
    the phase of the pattern near the peak response.  The histograms have
    two peaks which indicates that the pattern is mostly concentrated 
    near one of two phases; the two phases are separated by $\pi$.
    Forcing frequency: (a,d) $f_f=0.0500$~Hz, 
    (b,e) $f_f=0.0556$~Hz, 
    (c,f) $f_f=0.0625$~Hz.  
  }
  \label{fig:bz-standing}
\end{figure}

We differentiate the two types of response using their power spectra.  If
the system is resonant, the response frequency will adjust to be
a rational ratio of the forcing frequency.
For near-resonant response, however, the frequency does not 
adjust to be a rational ratio of the forcing frequency.
Figure~\ref{fig:bz-spiral} shows the power
spectra and corresponding histograms of phase angle for 2:1
near-resonant patterns (spirals) at three forcing frequencies near
exact resonance with the spiral wave frequency.  
The peak of the subharmonic response
does not adjust to $f_f/2$ (vertical line) as the frequency is varied.
The histograms of the phase, however, indicate that there is a two-phase
response to the forcing even though there is not exact resonance.  
In contrast, Fig.~\ref{fig:bz-standing} shows the power spectra of resonant
patterns (standing waves) for three forcing frequencies near exact resonance
with the uniform oscillation frequency.  In this case 
the patterns lock to $f_f/2$ (shown by the vertical line)
even when the forcing is detuned from exact resonance.

We now discuss the different types of $m\mbox{:}1$ patterns that we observed.

\subsection{1:1 and 2:1 patterns}
\label{sec:1to1}
\label{sec:2to1}
In the 1:1 region we observe a resonant response.
In this case the entire pattern of chemical concentration
oscillates uniformly in space with the forcing frequency, as measured for a
range of $f_f$ and $I$ values centered at $f_0$.
The shape of the 1:1 tongue in~Fig.~\ref{fig:alltongues}(a)
is different than the shape of the other tongues.  The other
tongues bend toward higher frequency at low forcing intensities.
Instead we find 1:1 uniform patterns at frequencies near $f_0$
even at very low forcing intensities
We do find spiral patterns at slightly higher frequencies
near the bottom of the tongue, but we cannot distinguish
1:1 spiral waves from unforced spiral waves.

Unlike the 1:1 resonant response, for which we observed only a single
qualitative pattern, several qualitatively different patterns were
observed inside of the 2:1 region.  In this region, the oscillation
phase responds to either the first or the second forcing cycle, which
occurs within a single oscillation cycle of the pattern.  The 2:1
patterns are therefore formed from spatial arrangements of regions
oscillating at the same frequency but which differ in phase by $\pi$.
A description of the different 2:1 patterns observed in the
BZ system and in a forced reaction-diffusion model with Brusselator
kinetics was given in~\cite{LBMS:00}.

\subsection{3:1 patterns}

\begin{figure}
  \includegraphics[width=3.25in]{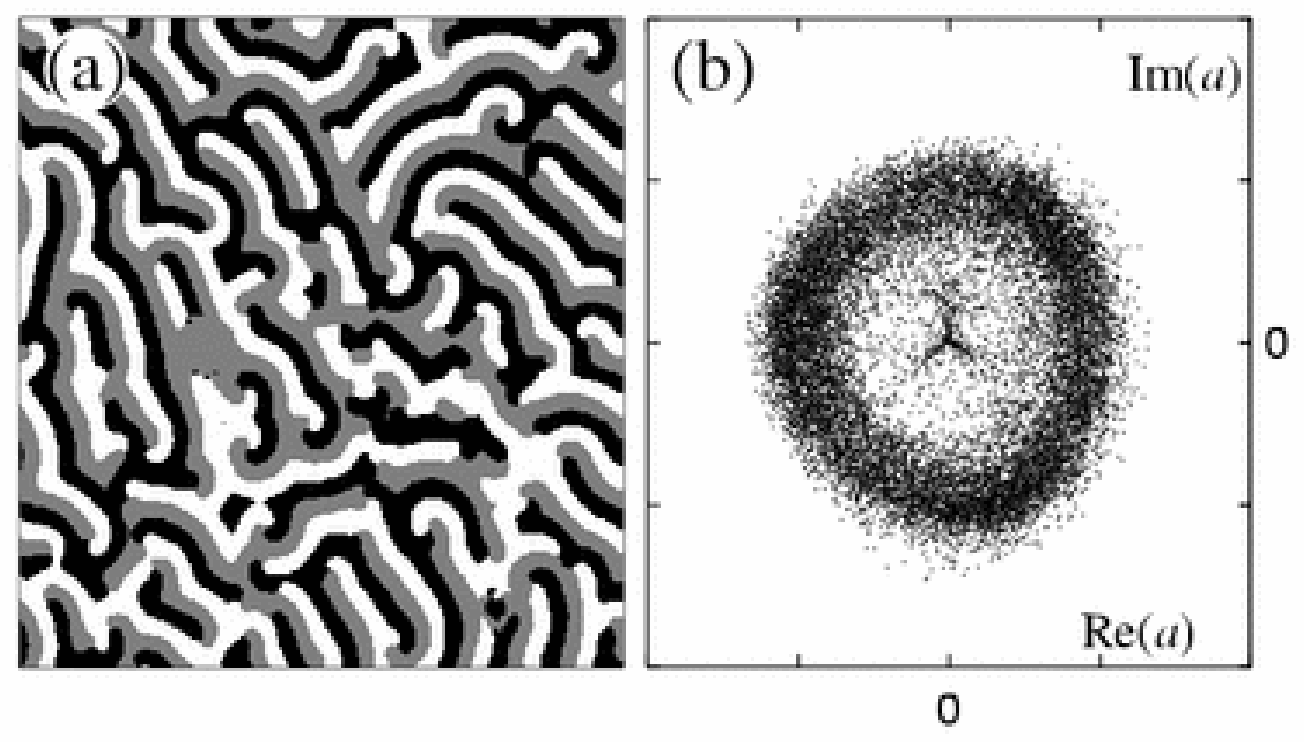}\\
  \vspace{-0.1in}
  \includegraphics[width=3.25in]{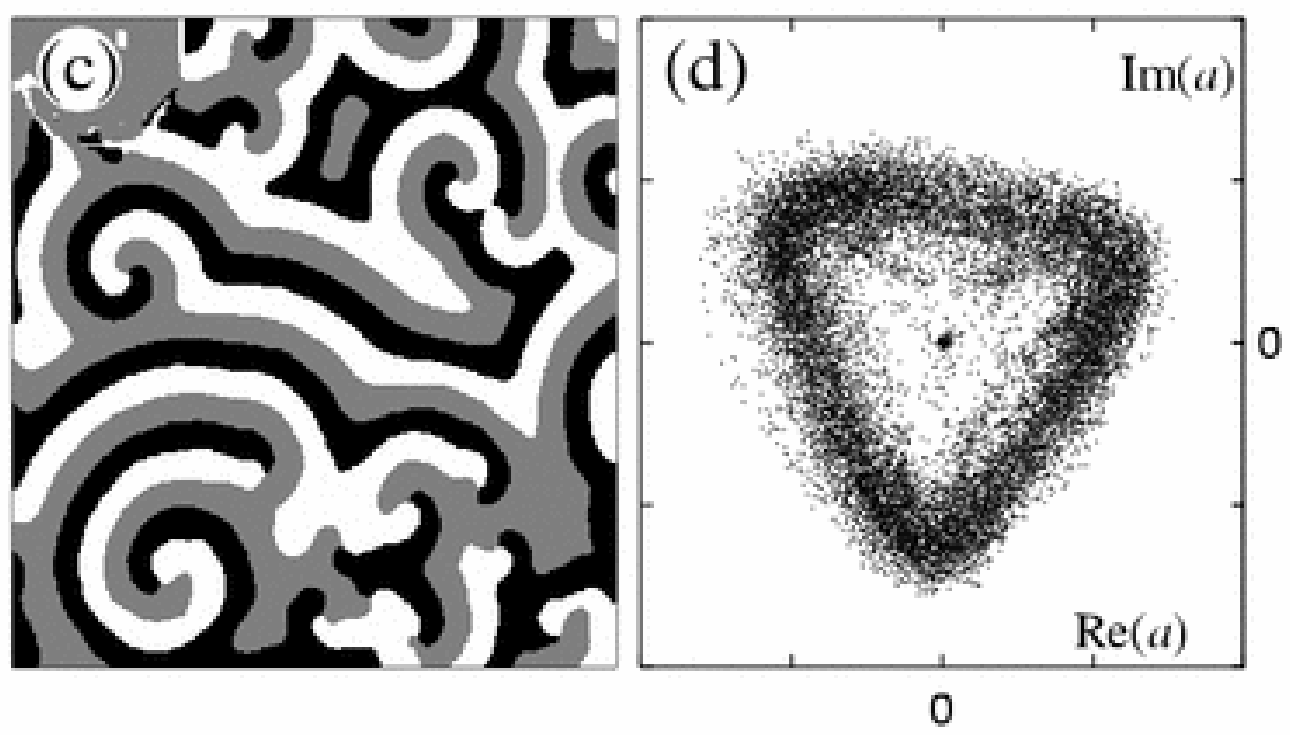}\\
  \vspace{-0.1in}
  \includegraphics[width=3.25in]{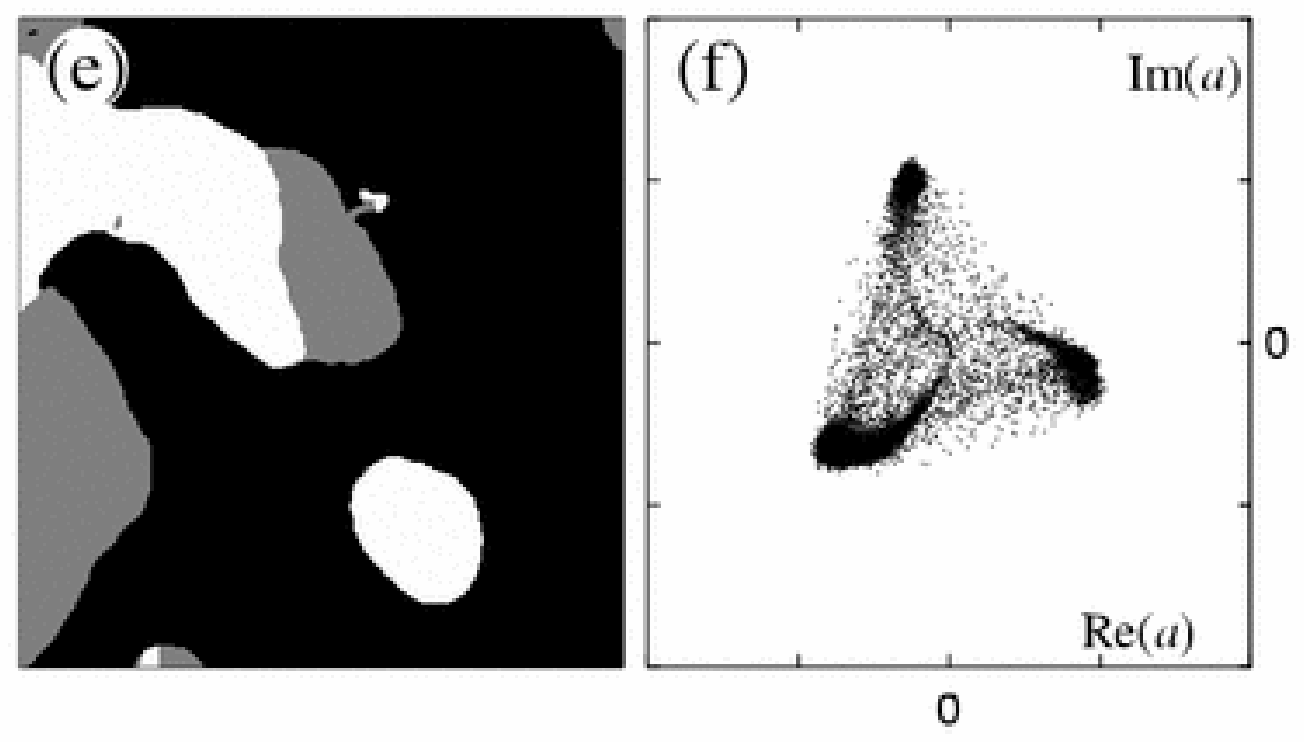}
  \vspace{-0.1in}
  \caption{
    Patterns observed in the 3:1 resonance region for different
    forcing amplitudes (the patterns are shown after filtering).  The
    response is shown in the $x-y$ plane (left) and complex phase plane
    (right).  (a,b) Spiral waves in the bottom of the 3:1 region, $I= 269
    \rm{W/m}^2$, $f_f = 0.1251$~Hz.  (c,d) Spiral waves in the middle
    of the 3:1 region, $I = 382 \rm{W/m}^2$, $f_f=0.0667$~Hz.  (d,e)
    Standing wave patterns found at the top of the 3:1 region, $I = 863
    \rm{W/m}^2$, $f_f=0.0769$~Hz.  
  }
  \label{fig:3to1snear_sfar_stand}
\end{figure}

In the 3:1 region we observe two qualitatively different types of
patterns.  At low forcing
the 3:1 patterns are rotating spirals, such as those shown in
Fig.~\ref{fig:3to1snear_sfar_stand}.  At low forcing intensity, the
spirals have a fairly evenly distributed phase angle and a
nearly circular shape in the complex phase plane. At higher forcing
intensity, the phase becomes more concentrated in three phases, and the
shape in the complex phase plane becomes more triangular.  This trend
is observable in Fig.~\ref{fig:3to1snear_sfar_stand}(b,c).

An abrupt transition from traveling wave patterns to what appear to be
standing wave patterns is observed in the 3:1 resonance region as the
forcing amplitude was increased.  The transition between these pattern
types was observed at a fixed forcing frequency of 0.075~Hz as $I$ was
increased past roughly 460 W/m$^2$, and was also observed for fixed
forcing amplitudes in a range of 300 to 400 W/m$^2$ when $f_f$ was
increased past roughly 0.065~Hz.  The experimental resolution is not
enough to determine the functional form of the transition.

The 3:1 standing wave patterns with stationary or nearly stationary 
fronts consist of irregularly shaped domains 
differing in phase by $2\pi/3$ (see Fig.~\ref{fig:3to1snear_sfar_stand}).  
Often the fronts
are rough, i.e., have short wavelength modulations that appear stable
over a hundred oscillation cycles of the pattern, as can be seen in
the standing wave pattern pictured in Fig.~\ref{fig:many}(c).

The fronts in the standing-wave patterns are either stationary or
propagate on a time scale orders of magnitude larger than the unforced
spiral period.  If the latter is the case, the patterns are not
precisely standing waves, and over a longer time could evolve to other
patterns such as large and slowly rotating three-phase spiral waves.
If so, these patterns could be understood to be observations of the 3:1
patterns predicted by the forced complex Ginzburg-Landau (CGL)
equation~\cite{GWST:01,CoEm:92}.

\subsection{Other patterns}
\label{sec:other}

Rotating four-phase spiral patterns, e.g. Fig.~\ref{fig:many}(d,h),
are the only pattern type observed in the 4:1 resonance region of the
BZ experiments.  A detailed description of 4:1 resonance in the forced
CGL equation and in the FitzHugh-Nagumo and Brusselator
reaction-diffusion models was given in~\cite{LBSA:00}.  In those
models a bifurcation from four-phase traveling patterns to two-phase
4:1 resonant standing-wave patterns was
found~\cite{lin00:_four_phase}.  This bifurcation has not been observed in
our experiments, perhaps because the forcing light intensity $I$
available may be insufficient to reach the
bifurcation.  Another possibility is that the patterns observed
were at a frequency near four times the spiral wave frequency $f_s$
instead of near four times the uniform oscillation frequency $f_0$.
In the experimentally accessible range of $I$, we have
observed only 4:1 spiral wave patterns and none of the more complicated
pattern behavior found in the 4:1 forced CGL model~\cite{Hemming_PHD}.

Patterns such as the 2:1 Bloch fronts and
spiral waves discussed in~\cite{CLHL:90,CoEm:92,EHMM:97} and the 4:1
rotating spirals discussed in~\cite{lin00:_four_phase} are not
resonant since they are all traveling patterns.

\begin{figure}
  \includegraphics[width=3.25truein]{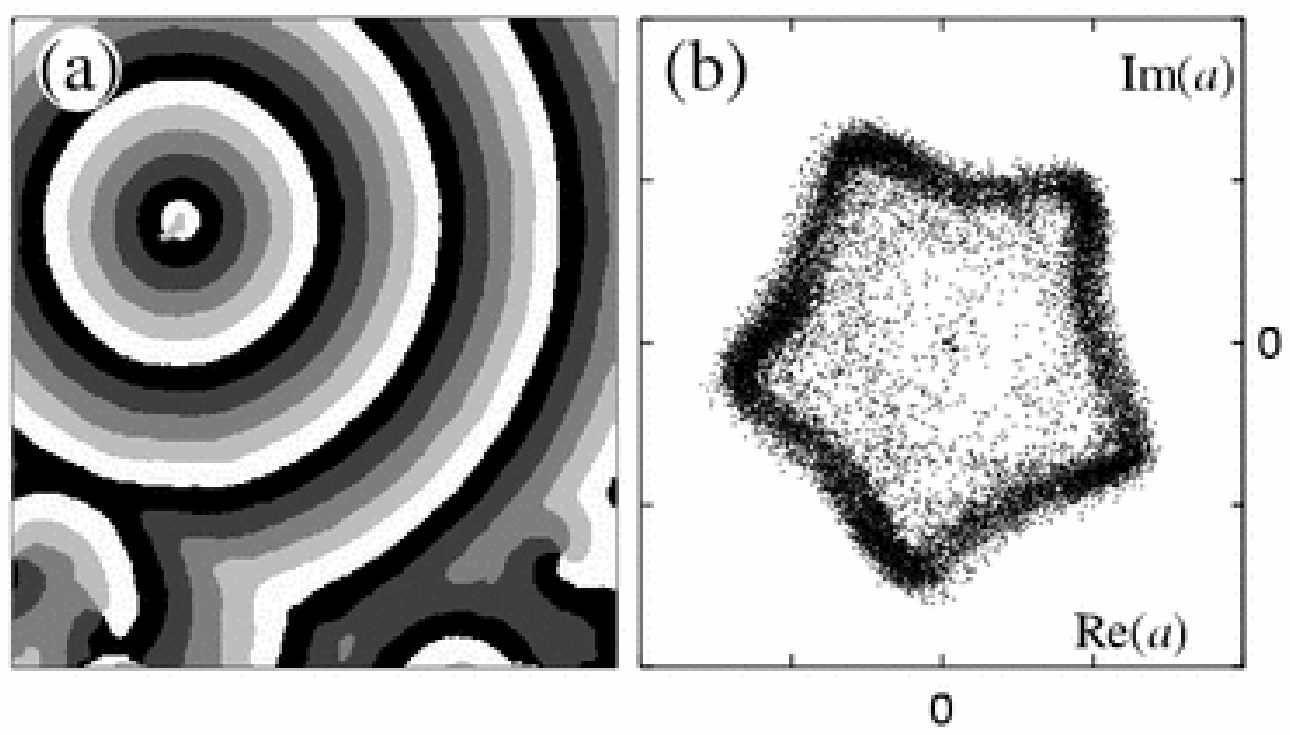}
  \includegraphics[width=1.7in]{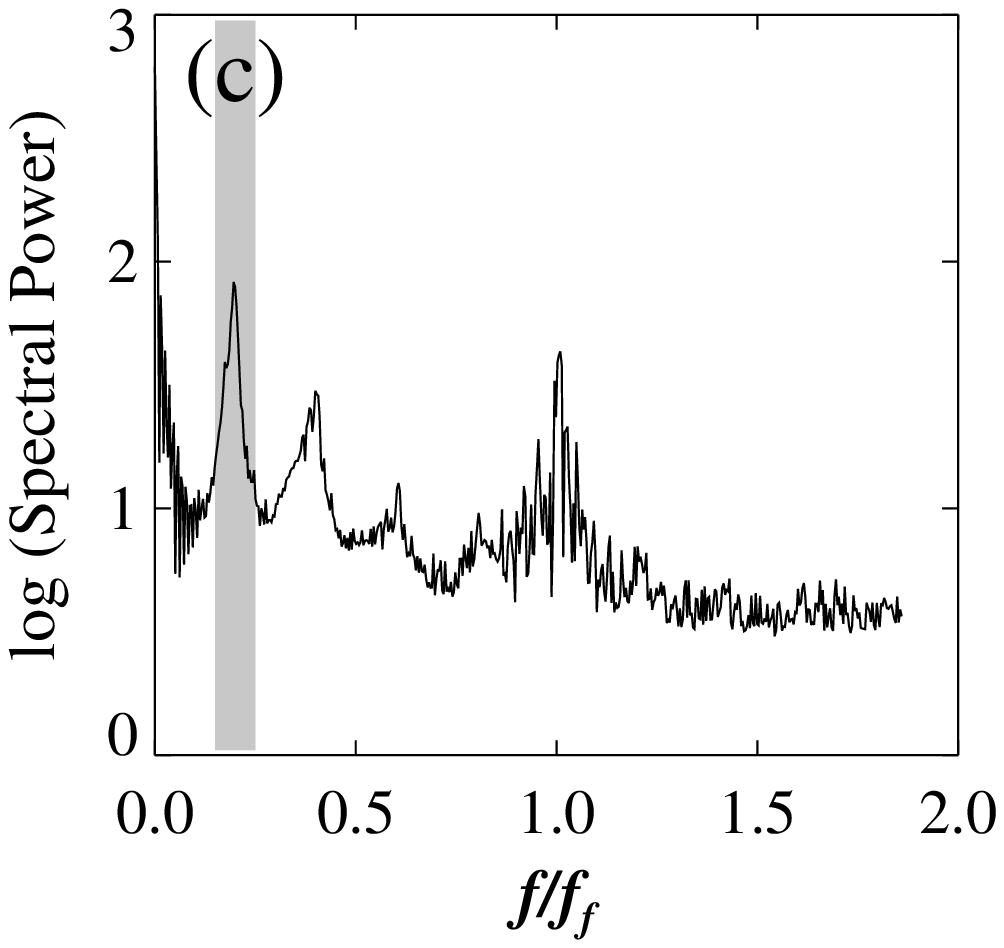}\includegraphics[width=1.7in]{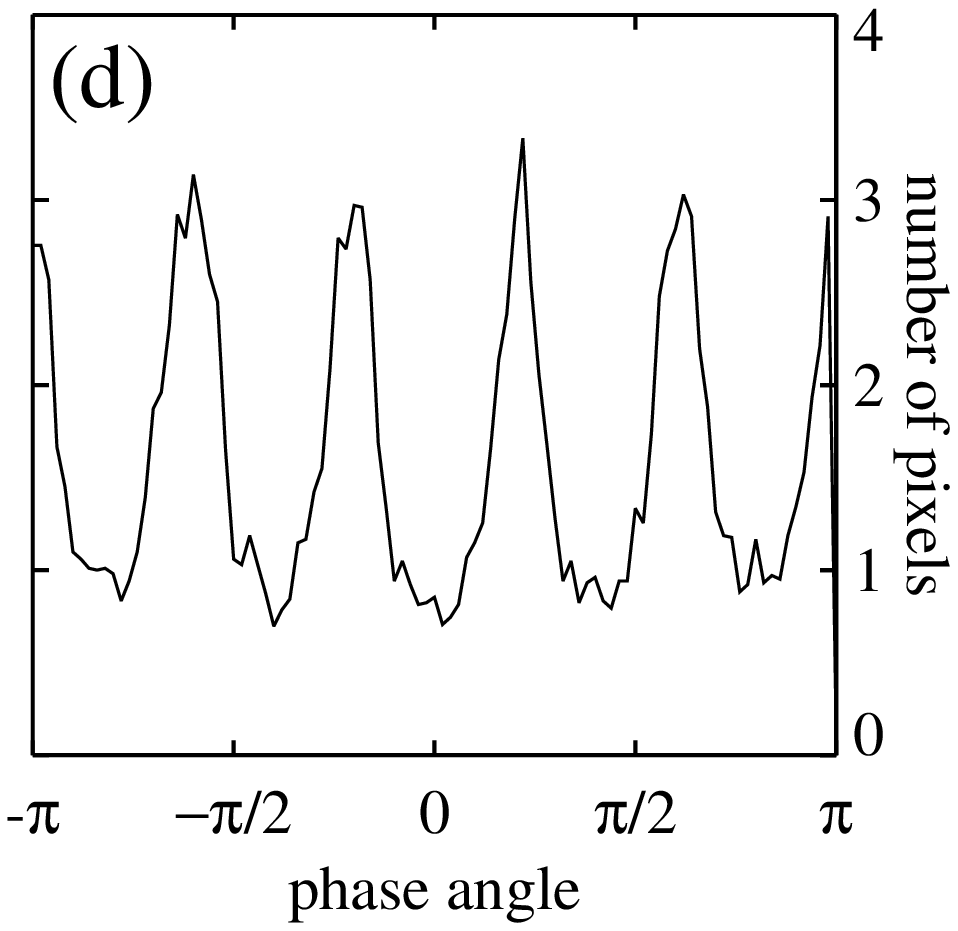}
  \caption{
    BZ spirals observed in the 5:1 resonance region
    at $I$ = 661 W/m$^2$, $f_f$ = 0.100~Hz.
    The patterns are filtered to keep only the frequencies shown
    by the gray band in (c).
    (a) Phase patterns in $x-y$ plane.
    (b) Phase in complex plane.
    (c) Average temporal power spectrum.
    (d) Histogram of phase angle.
  }
  \label{fig:5to1}
\end{figure}

\begin{figure}
  \includegraphics[width=3.25in]{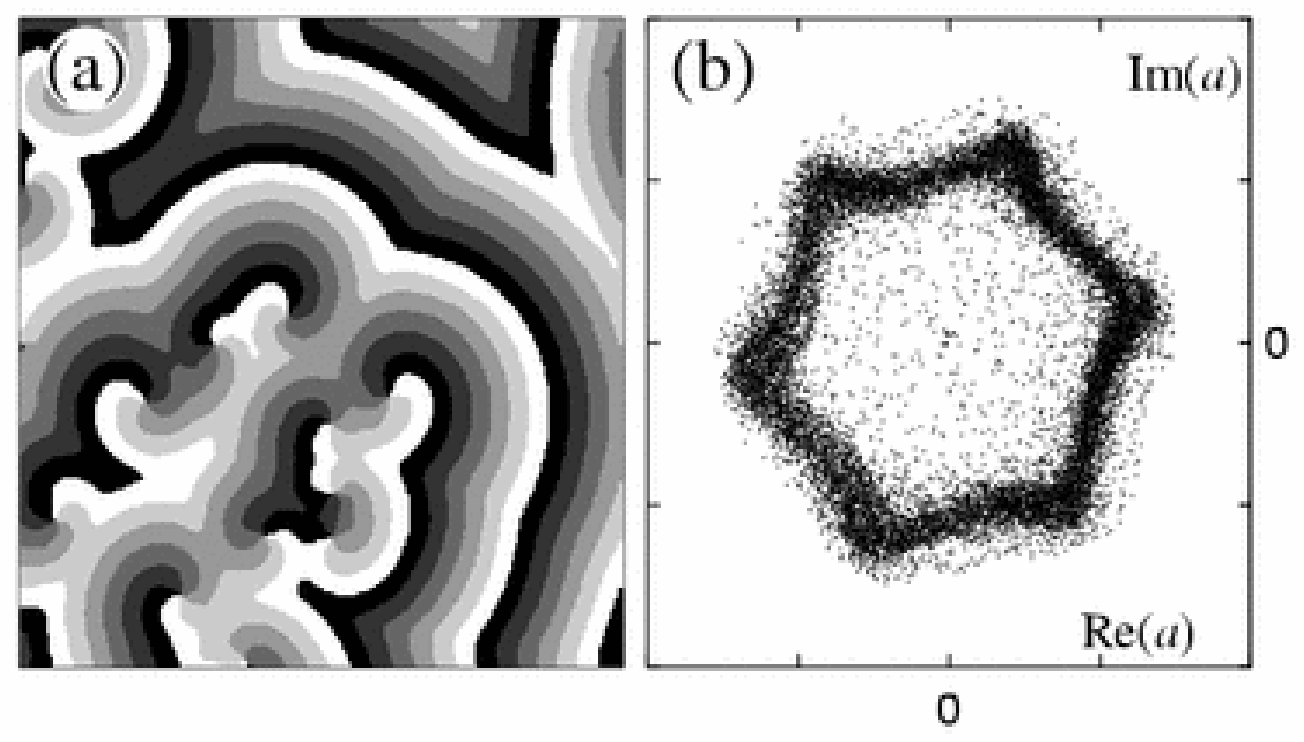}
  \includegraphics[width=1.7in]{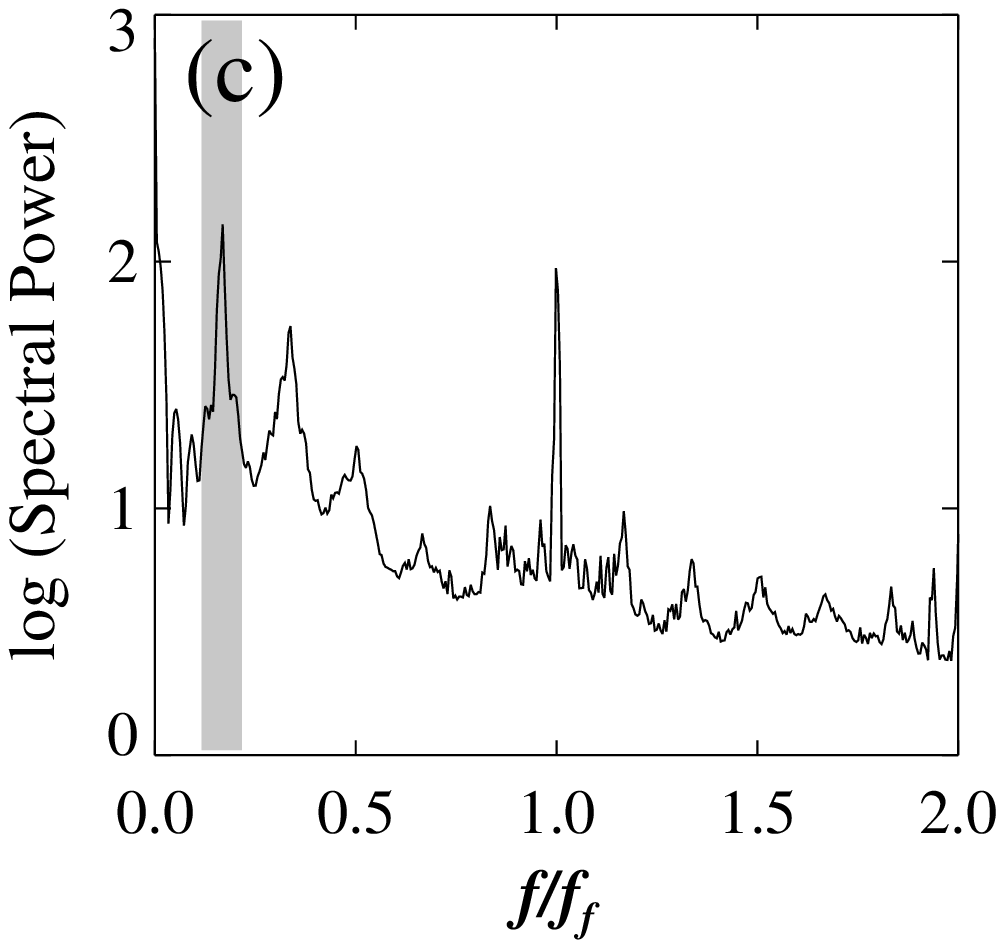}\includegraphics[width=1.7in]{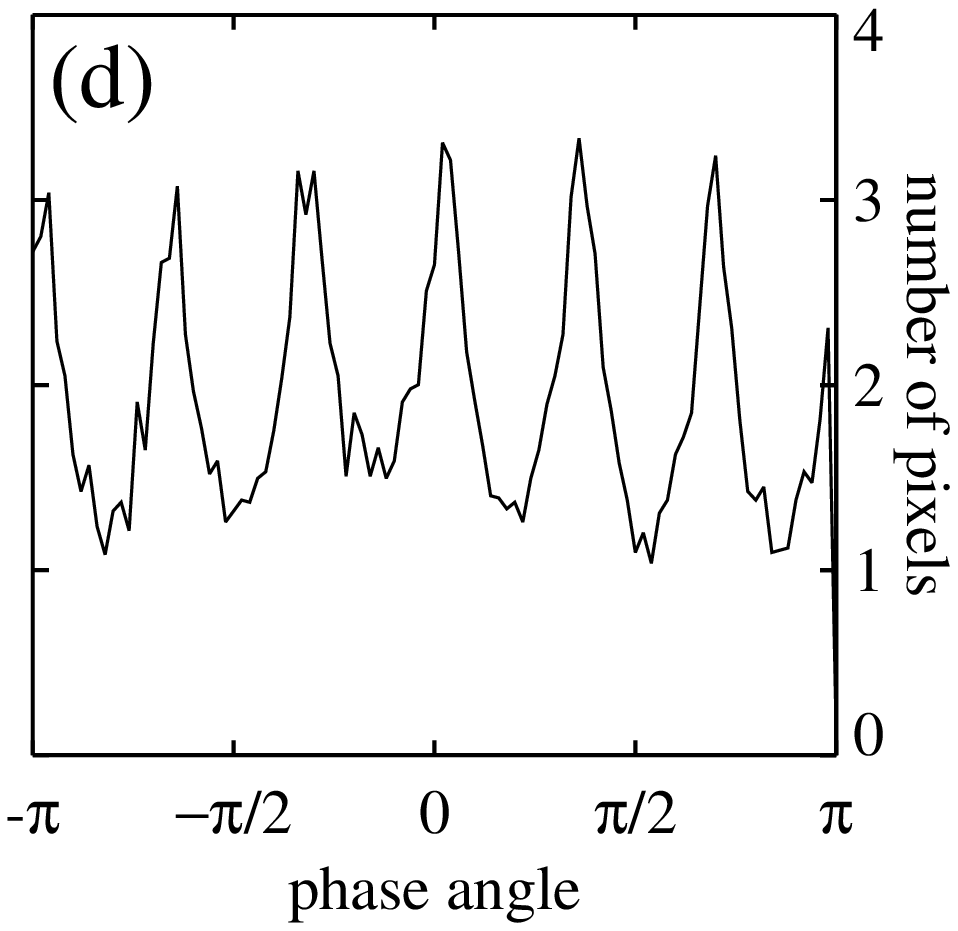}
  \caption{
    BZ spirals observed in the 6:1 resonance region
    at $I$ = 475 W/m$^2$, $f_f$ = 0.1001~Hz.
    The patterns are filtered to keep only the frequencies shown
    by the gray band in (c).
    (a) Phase patterns in the $x-y$ plane.
    (b) Phase in complex plane.
    (c) Average temporal power spectrum.
    (d) Histogram of phase angle.
  }
  \label{fig:6to1}
\end{figure}

Finally, we present examples of near resonant 5:1 and 6:1 patterns in
Fig.~\ref{fig:5to1} and Fig.~\ref{fig:6to1}, respectively.


\section{Reaction-diffusion model}
\label{sec:fhn}

As a model for a periodically forced oscillatory system we
use a version of the FitzHugh-Nagumo (FHN) reaction-diffusion equations
\begin{subequations}
\begin{eqnarray}
u_t &=& u - u^3 - v + \nabla^2 u \,,\\
v_t &=& \epsilon \left[ u - \left(a_1 + \gamma \sin\omega_f t\right) v \right] + \delta\nabla^2 v\,,
\end{eqnarray}
\label{eq:FHN}
\end{subequations}

\noindent
where the fields $u(x,y)$ and $v(x,y)$ represent concentrations
of chemicals in a simple model of a chemical system.
We add explicit time dependence to the FHN system as parametric
sinusoidal forcing with amplitude $\gamma$ and frequency $\omega_f$.
The parameter $\epsilon$ is the ratio of the time scales of $u$ and
$v$ and $\delta$ is the ratio of the diffusion rates of $u$ and $v$.
In the following
we fix the parameters $\epsilon=0.1$, $\delta=0.1$, $a_1=0.5$, and
vary the forcing frequency and amplitude.

In the absence of forcing, $\gamma=0$, the equations have a spatially
uniform solution $u=v=0$.  The parameter $\epsilon$  controls the
stability of this solution.  When $\epsilon>2$, $u=v=0$ is
stable, and at $\epsilon=2$, there is a Hopf bifurcation to uniform
oscillations.  Beyond the Hopf bifurcation Eqs.~(\ref{eq:FHN}) also support
traveling phase waves.  Our numerical investigations are conducted in
the parameter range where uniform oscillations and phase waves both exist.
In two space dimensions phase waves typically form into rotating spirals,
each one organized around a core where the amplitude of
oscillations is zero.  For the parameters above,
the spiral wave frequency ($\omega_s\approx 0.237$) is faster than
the homogeneous oscillation frequency ($\omega_s\approx 0.215$);
once formed, spiral waves spread to fill the entire system.

\subsection{Periodic forcing and data analysis}
\label{sec:fhnpatts}
A sinusoidal parametric forcing, homogeneous in space, is applied
by choosing a nonzero $\gamma$ parameter in Eq.~(\ref{eq:FHN}).
As in the BZ experiment, when the forcing amplitude is high
enough the system can lock at rational ratios of the forcing
frequency.  Frequency locking of spatially uniform
solutions to Eq.~(\ref{eq:FHN}) occurs in tongue-shaped
regions in the $\omega_f-\gamma$ parameter plane.
A complete diagram of the pattern-forming tongues was
not computed for the FHN equation with this forcing scheme.
We have only investigated certain resonances to compare
with the BZ chemical experiment.  Our numerical
investigations show that the size and shape of
different $m\mbox{:}n$ tongues depends on the exact form of the
parametric forcing in Eqs.~(\ref{eq:FHN})~\cite{bold03:_tongues}.
Tongue diagrams obtained by forcing other terms of the FHN model
were presented in studies of locking to uniform oscillations
in the oscillatory FHN model~\cite{Reza_MS,Xaver_MS}.
A diagram for the 2:1 tongue of a periodically forced Brusselator
reaction-diffusion system was given in Ref.~\cite{LBMS:00}.

The data from the numerical solutions of the forced FHN equation are
processed to extract phase information, as in the experimental system.
Every 1.4 time units, we store the values of $u(x,y)$ and $v(x,y)$ at
each computational grid point $x_i, y_j$.  The discrete Fourier
transform is applied to the time variable of $u(x,y,t)$ to get the
frequency response $\hat{u}(x,y,\omega)$ for each point in the
pattern.  The averaged power spectrum of the signal,
\begin{equation}
P(\omega)=\frac{1}{N_x N_y}\sum_{i,j} |{\hat u(x_i,x_j)}|^2\,,
\end{equation}
where $N_x$ and $N_y$ are the number of grid points in the $x$ and $y$
directions, is then examined to determine the system response.
The response frequency is isolated from the signal $\hat{u}$
using a box filter centered at the response frequency,
$\omega_r$ with width $\Delta$.
The filtered signal is then inverse Fourier transformed, which gives
the response in time, $a(x,y,t)$.
\begin{figure}
  \includegraphics[width=3.25in]{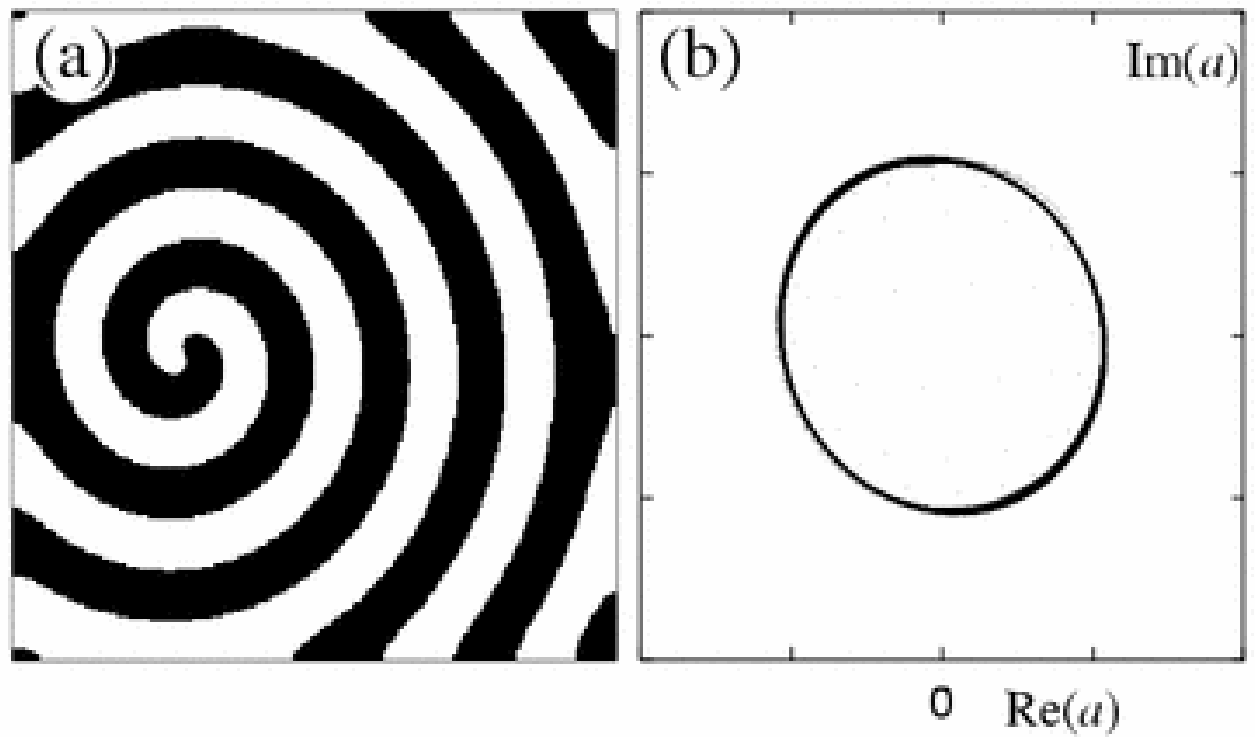}
  \includegraphics[width=1.7in]{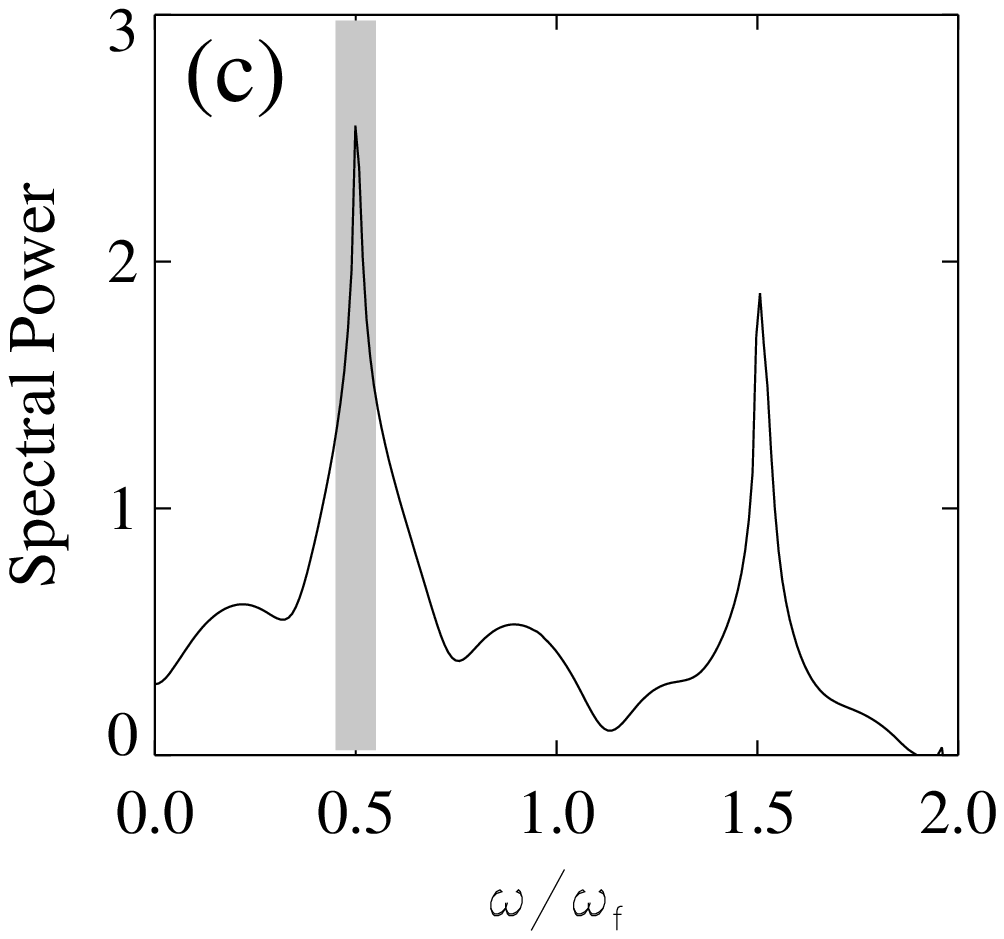}\includegraphics[width=1.7in]{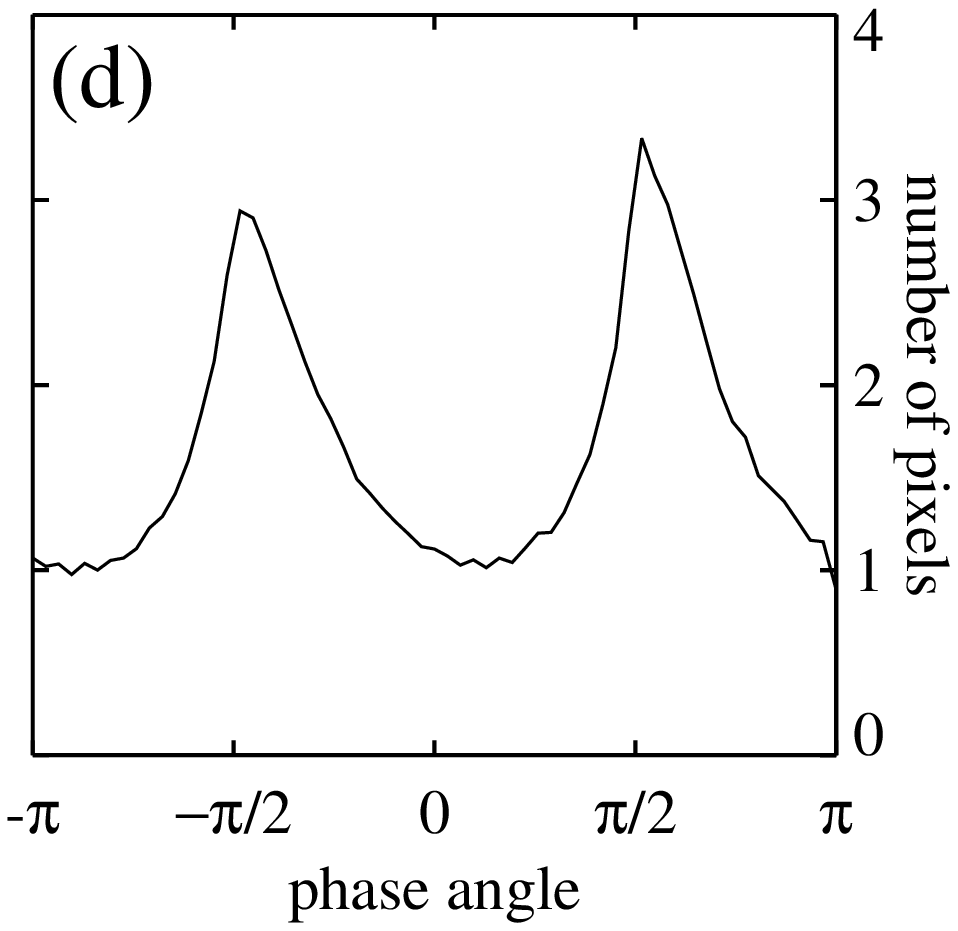}
  \caption{
    Spiral wave in Eq.~(\protect\ref{eq:FHN}) with periodic forcing at
    nearly twice the spiral wave frequency.
    The patterns are filtered to keep only the frequencies shown
    by the gray band in (c).
    (a) Phase pattern in $x-y$ plane.
    (b) Phase in complex plane.
    (c) Average temporal power spectrum.
    (d) Histogram of phase angle.
    Parameters: $\gamma=0.5, \omega_f=0.474$, $x=[0,256]$, $y=[0,256]$.
  }
  \label{fig:FHN2}
\end{figure}

Figure~\ref{fig:FHN2} shows an example of a 2:1 spiral wave with the
peak frequency response at $\omega_f/2$ displayed clearly in the power
spectra.  The filtered signal $a$ in $x-y$ domain is shown in
Fig.~\ref{fig:FHN2}(a), and Fig.~\ref{fig:FHN2}(b) shows the same data
plotted in the complex phase plane.  The width of the filter is shown
in the power spectrum by the gray band (see Fig.~\ref{fig:FHN2}(c)).
The phase plane shows that different parts of the spatial domain are
in different relative phases, all oscillating at the same frequency.
The phase is not uniformly distributed but has peaks near two phases
that become apparent in the histogram of the phase angle shown in
Fig.~\ref{fig:FHN2}(d).

\subsection{Pattern Formation}
The FHN equations~(\ref{eq:FHN}) have two intrinsic frequencies, the
uniform oscillation frequency, $\omega_0$, and the spiral wave
frequency $\omega_s$.  For some choices of parameters
(e.g. $\delta=1$) these two frequencies are the same, but for the
parameters chosen in this study the two frequencies differ ($\omega_s
> \omega_0$).  Because of this there are two possible different
resonant response conditions: when the forcing is a rational multiple
of either $\omega_0$ or $\omega_s$.  In the following we will show how
the FHN equations~(\ref{eq:FHN}) respond to forcing in both of those
cases.  In the BZ experiment this distinction is harder to make.

\subsection{Patterns at $m\mbox{:}n$ response of the uniform oscillation frequency}
When the forcing is a rational multiple of the {\sl uniform} oscillation
frequency $\omega_0$, spatially uniform solutions of Eq.~(\ref{eq:FHN})
are found for a range of forcing frequency $\omega_f$ and amplitude
$\gamma$.  The $m\mbox{:}n$ frequency-locked solutions form tongue shaped
regions in the $\omega_f-\gamma$ parameter plane (Arnol'd tongues).
The shape of the resonant tongues depends on
the exact form of the forcing~\cite{bold03:_tongues}.

As in the BZ experiment, patterns may form in the frequency-locked
tongues.  Resonant pattern solutions consist of standing
waves connecting regions of different phases.  For example,
in the 2:1 resonance, standing waves consist of fronts between
regions in space that are oscillating at the same frequency
but out of phase by $\pi$.  The fronts must be stationary
for the pattern to be strictly frequency locked, since any motion
indicates that the phase is drifting and thus, at least in the vicinity
of a front, the frequency is also slowly changing.

A 2:1 resonance is found when the forcing frequency is nearly twice
the uniform oscillation frequency, $\omega_f\sim2\omega_0$.  For
sufficiently high forcing amplitude, the system frequency locks at
$\omega_f/2$, in one of two phases separated by $\pi$.  Patterns form
from the two phases.  At low forcing amplitude the pattern is a two
phase rotating spiral wave and thus is near-resonant but not strictly
frequency-locked since the spiral rotates slowly (relative to the
forcing period).  At higher forcing amplitude the pattern is standing
waves.  The spiral waves and
standing waves are similar to those found in the BZ
experiment and in forced complex Ginzburg-Landau and Brusselator
models~\cite{POS:97,Bertram_MS,LBMS:00,yochelis02:_devel}.

In the 3:1 resonance the system responds at one-third the forcing
frequency, $\omega_r=\omega_f/3$, and the patterns consist of spatial
regions of three locked phases.  In our exploration in forced FHN
model we find that the three locked phases organize into rotating
spiral waves.

Patterns in a forced 4:1 FHN model consist of four-phase spiral waves
and two-phase standing waves and are discussed in detail in
Ref.~\cite{lin00:_four_phase}.

The domain in parameter space where frequency locked patterns
(standing waves) exist is different from that of frequency locked
uniform solutions.  Recently it was discovered that frequency locked
standing wave patterns can exist outside of the resonant tongue of
spatially uniform solutions or that spatial instabilities can reduce
the range of resonant
patterns~\cite{yochelis02:_standing,yochelis03:_impact}.

\subsection{Patterns at $m\mbox{:}n$ response of the spiral wave frequency}

\begin{figure}
  \includegraphics[width=3.0in]{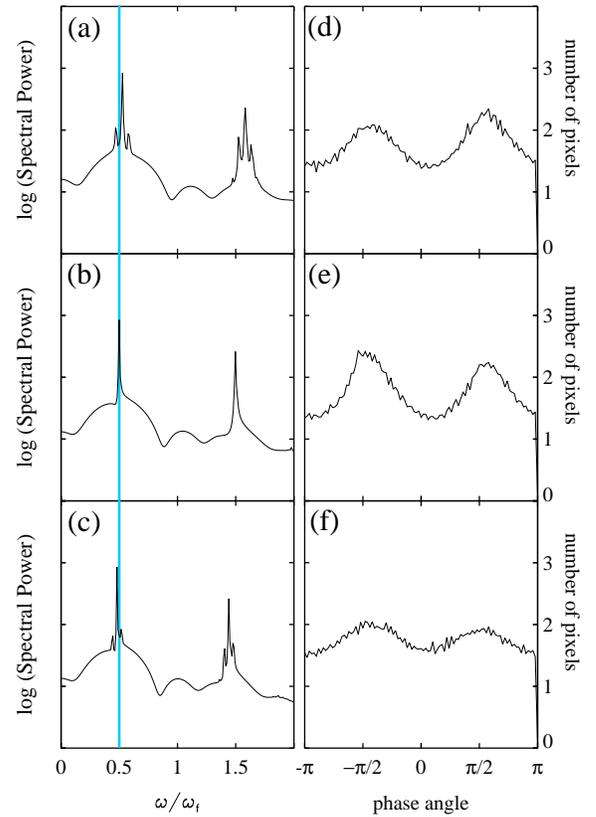}
  \caption{
    Response of a 2:1 forced spiral wave
    in the FHN Eq.~(\protect\ref{eq:FHN}).
    (a)-(c) Power spectra for three different forcing frequencies
    (a) $\omega_f/2 < \omega_s$,
    (b) $\omega_f/2 \approx \omega_s$,
    (c) $\omega_f/2 > \omega_s$.
    The solid line indicates $\omega_f/2$.
    When the forcing is not exactly in resonance the peak response
    in the power spectra differs from $\omega_f/2$ indicating
    that the spiral is not frequency locked.
    (d)-(e) Histograms of the distribution of phase angle for the
    spiral waves in (a)-(c).  The two peaks in the distribution
    show that the spiral wave responds to the forcing by
    redistributing the internal relative distribution of the phase
    even though the pattern is not frequency locked.
    Parameters: $\gamma=0.3$,
    (a) $\omega_f=0.465$,
    (b) $\omega_f=0.474$,
    (c) $\omega_f=0.493$.
  }
  \label{fig:fhn-spiral}
\end{figure}

When the forcing frequency is close to a rational multiple of the spiral wave
frequency, the spiral does not frequency lock to the forcing but still
shows a near-resonant response.  This can be seen in the nonuniform
phase distribution of the forced spiral waves, with $m$ peaks for waves
forced near an $m\mbox{:}n$ resonance.  For example, forcing the spiral wave at
approximately twice the spiral frequency, $\omega_f\approx2\omega_s$,
causes the spiral to respond by shifting the relative oscillation
phases within the spiral to be concentrated near two phases,
as shown in Figures~\ref{fig:fhn-spiral}(d)-(f).

Figure~\ref{fig:fhn-spiral} shows the spiral wave response when the
forcing frequency is scanned through the spiral frequency.  The
maximum response in the power spectrum is not $\omega_f/2$ when the
forcing frequency is not exactly twice the spiral wave frequency; the
pattern is not frequency locked but it is near-resonant.  In
contrast to a quasi-periodic response farther away from
resonance~\cite{LBMS:00}, near-resonant patterns exhibit a nonuniform
distribution in the histograms of the phase angle.  This distribution
is farthest from uniform when the forcing is closest to $\omega_f/2$,
see Fig.~\ref{fig:fhn-spiral}(d)-(f).
\begin{figure}
  \includegraphics[width=3.0in]{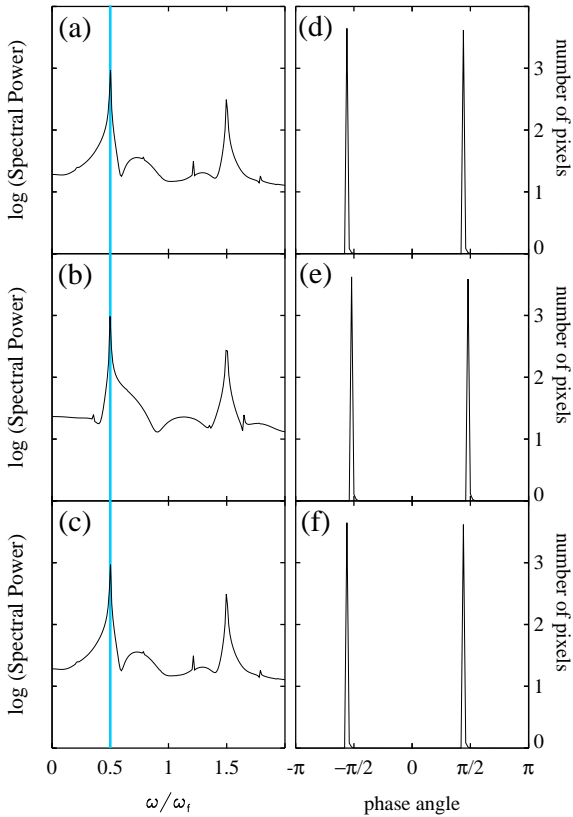}
  \caption{
    Resonant standing-wave response in the 2:1 forced
    FHN Eq.~(\protect\ref{eq:FHN}).
    (a)-(c) Power spectra for three different forcing frequencies
    near $2\omega_0$.
    The frequencies are normalized to the forcing frequency $\omega_f$.
    The peak
    subharmonic response is exactly at half the forcing frequency,
    even when the forcing is not exactly $2\omega_0$,
    as indicated by the vertical line at  $\omega/\omega_f=0.5$.
    (d)-(e) Histograms of the distribution of phase angle for the
    spiral waves in (a)-(c).  The two peaks in the distribution
    show that the pattern response is primarily in two phases
    separated by an angle of $\pi$.
    Parameters: $\gamma=3.0$,
    (a) $\omega_f=0.349$,
    (b) $\omega_f=0.370$,
    (c) $\omega_f=0.419$.
  }
  \label{fig:fhn-standing}
\end{figure}

Forcing near other resonances of the spiral wave frequency also
results in a near-resonant response, with the number of peaks in the
phase histogram corresponding to the $m\mbox{:}n$ resonance.  In the 3:1
resonance we observe three peaks and in the 4:1 resonance we find four
peaks.  Other spiral resonances such as 5:1 and 6:1 can also be found.
These results are similar to observations near-resonant spirals in the
of the BZ experiment.

To characterize the effect of the forcing on the spiral wave pattern,
we measured the deviation of the phase from a uniform distribution.
For an unforced spiral wave, the histogram of the phase
near the spiral frequency is flat, indicating that the
phase is uniformly distributed between $-\pi$ and $\pi$.  When
the spiral wave is nearly 2:1 resonant, the histogram shows two
peaks that are separated by $\pi$ in the phase distribution.
Figures~\ref{fig:fhn-spiral}(d)-(f) show.

\begin{figure}
  \centering
  \includegraphics[width=3.0in]{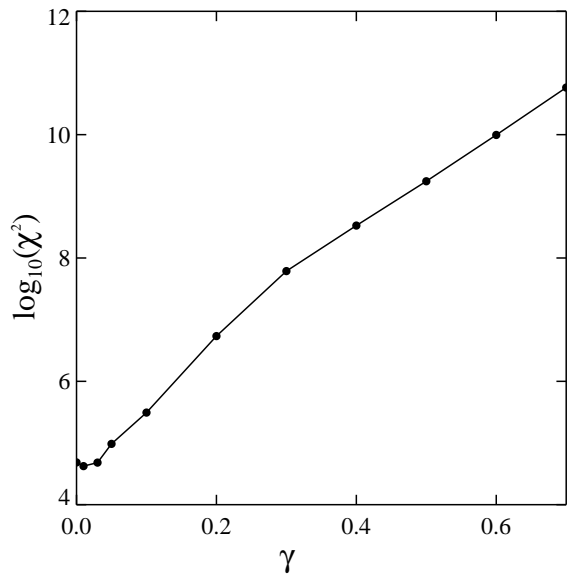}
  \caption{
    The $\chi^2$ statistic of the deviation in the
    phase distribution of the 2:1 forced spiral wave from a uniform
    (unforced spiral) distribution.  The $\chi^2$ value increases
    exponentially as the forcing amplitude increases.  Parameters in
    Eq.~\protect\ref{eq:FHN}: 
    $\epsilon=0.1, \delta=0.1, a_1=0.5, \omega_f=0.465$. 
  }
  \label{fig:sdev}
\end{figure}

The nonuniform phase response was measured by the chi-square statistic
relative to the uniform distribution.  The phase data at
$k=256\times256$ computational grid points was binned into 100 equal
size bins between $-\pi$ and $\pi$.  The chi-square statistic is

\begin{equation}
\chi^2 = \sum_i \frac{(N_i-E)^2}{E}\,,
\end{equation}
where $N_i$ is the value in bin $i$ and $E=k/100$ is the expected
value. Figure~\ref{fig:sdev} shows the dependence of $\chi^2$ on the
forcing amplitude $\gamma$ when the forcing frequency is near
$2\omega_s$.  The $\chi^2$ value increases exponentially as the
forcing amplitude is increased from near zero.  The nonzero value of
$\chi^2$ at $\gamma=0$ is due to small fluctuations in the phase of
our finite size sample.

\section{Discussion}
\label{sec:discussion}

The BZ chemical system in an open gel-reactor was used to identify
multiple tongues, each with a different $m\mbox{:}n$ resonance, in the forcing
frequency-amplitude parameter plane.  Such a phase diagram has not
been previously reported for a spatially extended oscillatory system.  The
resonance tongues are found to be ordered in a Farey sequence, similar
to the Devil's staircase ordering of resonance tongues for two coupled
oscillators~\cite{PBak:86} and for the homogeneous BZ
reaction~\cite{maselko86, maselko87}. 

The diffusively coupled oscillations we measure respond to external
forcing either resonantly or at near resonance (quasi-periodically but
with an $m$-peaked phase distribution).  The resonant patterns are
standing waves that frequency lock to a $m\mbox{:}n$ ratio of the
forcing frequency.  In this case, a power spectrum of the resulting
resonant pattern shows a single primary peak at $f_f/m$, along with
its higher harmonics, and the phase distribution has $m$ peaks shifted
by $2\pi/m$.  The near-resonant patterns are traveling waves which do
not lock to a ratio of the forcing frequency but have a response {\sl
near} $f_f/m$.  However, the phase distribution still shows $m$ peaks.
This near-resonant quasi-periodic behavior is different from
quasi-periodicity farther away from resonance, where patterns have a
flat phase distribution~\cite{LBMS:00,Bertram_MS}.

Both the resonant and near-resonant behavior are also
observed in a FitzHugh-Nagumo reaction-diffusion model
with sinusoidal periodic forcing, similar to the experiments.

\acknowledgments
We thank Karl Martinez for help with the experiment. The experiments were
conducted at the University of Texas at Austin with the support of the
Robert A. Welch Foundation and the Engineering Research Program of the
Office of Basic Energy Sciences of the Department of Energy.
The present collaboration was supported by Grant No. 98-0129 from the
United States - Israel Binational Science Foundation,
the Department of Energy under contracts W-7405-ENG-36,
and the DOE Office of Science Advanced Computing Research program in
Applied Mathematical Sciences.

\bibliography{multiphase}

\begin{thebibliography}{44}
\bibitem[{BZ1()}]{BZ1}
\bibinfo{note}{The chemical concentrations in the two chemical reserviors
  separated by the porous membrane were, Reservior I: $0.220$ M bromo-malonic
  acid, $0.230$ M potassium bromate, $0.80$ M sulfuric acid; Reservior II:
  $0.184$ M potassium bromate, $1\times10^{-3}$ M
  Tris(2,2'-bipyridyl)dichlororuthenium(II)hexahydrate, $0.80$ M sulfuric acid.
  Each reservior volume is 8.3 ml and the flow rate of chemicals through
  Reservior I was 20 ml/hr while through Reservior II it was 5 ml/hr. Chemicals
  were premixed before entering each reservior; a 10 ml premixer and a 0.5 ml
  premixer fed Reservior I and II, respectively. The experiments were conducted
  at room temperature.}
\bibitem[{BZ2()}]{BZ2}
\bibinfo{note}{The chemical concentrations in the two chemical reserviors are
  the same as reported in \protect\cite{BZ1}, except $0.300$ M bromo-malonic
  acid and $0.0136$ M potassium bromate were used.}
\end{thebibliography}

\end{document}